\documentclass[prl,twocolumn,showpacs,rotating,preprintnumbers,amsmath,amssymb]{revtex4}
\usepackage{multirow}
\usepackage{graphicx}
\usepackage{dcolumn,color}
\usepackage{bm}
\usepackage{epsfig}

\begin{document}
\bibliographystyle{try}
\newcommand{\er}{$\pm$}
\newcommand{\be}{\begin{eqnarray}}
\newcommand{\ee}{\end{eqnarray}}
\newcommand{\widt}{\rm\Gamma_{tot}}
\newcommand{\wadd}{$\rm\Gamma_{miss}$}
\newcommand{\gpiN}{$\rm\Gamma_{\pi N}$}
\newcommand{\getN}{$\rm\Gamma_{\eta N}$}
\newcommand{\gkla}{$\rm\Gamma_{K \Lambda}$}
\newcommand{\gksi}{$\rm\Gamma_{K \Sigma}$}
\newcommand{\gNpi}{$\rm\Gamma_{P_{11} \pi}$}
\newcommand{\gDpf}{$\rm\Gamma_{\Delta\pi(L\!<\!J)}$}
\newcommand{\gDps}{$\rm\Gamma_{\Delta\pi(L\!>\!J)}$}
\newcommand{\sqgDpf}{$\rm\sqrt\Gamma_{\Delta\pi(L\!<\!J)}$}
\newcommand{\sqgDps}{$\rm\sqrt\Gamma_{\Delta\pi(L\!>\!J)}$}
\newcommand{\gnsi}{$\rm N\sigma$}
\newcommand{\gNpf}{$\rm\Gamma_{D_{13}\pi}$}
\newcommand{\gNps}{$\Gamma_{D_{13}\pi(L\!>\!J)}$}
\newcommand{\roper}{$\rm N(1440)P_{11}$}
\newcommand{\srma}{$\rm N(1535)S_{11}$}
\newcommand{\trma}{$\rm N(1520)D_{13}$}
\newcommand{\srmb}{$\rm N(1650)S_{11}$}
\newcommand{\trmb}{$\rm N(1700)D_{13}$}
\newcommand{\trmc}{$\rm N(1875)D_{13}$}
\newcommand{\trmd}{$\rm N(2170)D_{13}$}
\newcommand{\fvma}{$\rm N(1675)D_{15}$}
\newcommand{\fvmb}{$\rm N(2070)D_{15}$}
\newcommand{\fvpa}{$\rm N(1680)F_{15}$}
\newcommand{\srpb}{$\rm N(1710)P_{11}$}
\newcommand{\trpa}{$\rm N(1720)P_{13}$}
\newcommand{\trpb}{$\rm N(2200)P_{13}$}
\newcommand{\trpd}{$\rm N(2170)D_{13}$}
\newcommand{\dtpa}{$\rm\Delta(1232)P_{33}$}
\newcommand{\doma}{$\rm\Delta(1620)S_{31}$}
\newcommand{\dtma}{$\rm\Delta(1700)D_{33}$}
\newcommand{\dtmb}{$\rm\Delta(1940)D_{33}$}
\newcommand{\rthe}{A^{1/2}/A^{3/2}}
\newcommand{\amoh}{$A^{1/2}$}
\newcommand{\amth}{$A^{3/2}$}
\newcommand{\broh}{\Gamma^{1/2}_{\gamma p}/\widt}
\newcommand{\brth}{\Gamma^{3/2}_{\gamma p}/\widt}
\newcommand{\btot}{$\Gamma(\gamma p)$}
\newcommand{\Dpi}{\Delta\pi}
\newcommand{\KL}{\rm\Lambda K}
\newcommand{\KS}{\rm\Sigma K}

\newcounter{univ_counter}
\setcounter{univ_counter} {0} \addtocounter{univ_counter} {1}
\edef\HISKP{$^{\arabic{univ_counter}}$ }
\addtocounter{univ_counter}{1}
\edef\GATCHINA{$^{\arabic{univ_counter}}$ }
\addtocounter{univ_counter}{1}
\edef\ERLANGEN{$^{\arabic{univ_counter}}$ }
\addtocounter{univ_counter}{1} \edef\PI{$^{\arabic{univ_counter}}$ }
\addtocounter{univ_counter}{1} \edef\KVI{$^{\arabic{univ_counter}}$}
\addtocounter{univ_counter}{1} \edef\FSU{$^{\arabic{univ_counter}}$
} \addtocounter{univ_counter}{1}
\edef\BOCHUM{$^{\arabic{univ_counter}}$ }
\addtocounter{univ_counter}{1}
\edef\BASEL{$^{\arabic{univ_counter}}$ }
\addtocounter{univ_counter}{1}
\edef\GIESSEN{$^{\arabic{univ_counter}}$ }
\addtocounter{univ_counter}{1}

\title{Evidence for a parity doublet \boldmath$\Delta(1920)P_{33}$ and \boldmath$\Delta(1940)D_{33}$\unboldmath\ from
\boldmath$\gamma p\to p\pi^0\eta$\unboldmath \\
\vskip 5mm }

\author{
I.\,Horn\,\mbox{\HISKP\hspace{-0.5mm},}
A.V.~Anisovich\HISKP\hspace{-1mm}$^,$\GATCHINA\hspace{-1mm},
G.~Anton\ERLANGEN\hspace{-1mm}, R.~Bantes\PI\hspace{-1mm},
O.~Bartholomy\HISKP\hspace{-1mm}, R.~Beck\HISKP\hspace{-1mm},
Yu.~Beloglazov\mbox{\GATCHINA\hspace{-1mm},}
R.~Castelijns\KVI\hspace{-1mm}, V.~Crede\FSU\hspace{-1mm},
A.~Ehmanns\mbox{\HISKP\hspace{-1mm},}
J.~Ernst\HISKP\hspace{-1mm}, I. Fabry\HISKP\hspace{-1mm},
H.~Flemming\BOCHUM\hspace{-1mm}, A.~F\"osel\ERLANGEN\hspace{-1mm},
M.~Fuchs\mbox{\HISKP\hspace{-1mm},}
Chr.~Funke\mbox{\HISKP\hspace{-1mm},} R.~Gothe\PI\hspace{-1mm},
A.~Gridnev\mbox{\GATCHINA\hspace{-1mm},} E.~Gutz\HISKP\hspace{-1mm},
St.~H\"offgen\PI\hspace{-1mm}, J.~H\"o\ss
l\mbox{\ERLANGEN\hspace{-1mm},} J.~Junkersfeld\HISKP\hspace{-1mm},
H.~Kalinowsky\mbox{\HISKP\hspace{-1mm},}
F.~Klein\mbox{\PI\hspace{-1mm},} E.~Klempt\HISKP\hspace{-1mm},
H.~Koch\BOCHUM\hspace{-1mm}, M.~Konrad\mbox{\PI\hspace{-1mm},}
B.~Kopf\mbox{\BOCHUM\hspace{-1mm},}
B.~Krusche\mbox{\BASEL\hspace{-1mm},}
J.~Langheinrich\mbox{\PI\hspace{-1mm},}
H.~L\"ohner\KVI\hspace{-1mm},
I.~Lopatin\mbox{\GATCHINA\hspace{-1mm},} J.~Lotz\HISKP\hspace{-1mm},
H.~Matth\"ay\BOCHUM\hspace{-1mm}, D.~Menze\mbox{\PI\hspace{-1mm},}
J.~Messchendorp\GIESSEN\hspace{-1mm}, V.~Metag\GIESSEN\hspace{-1mm},
V.A.~Nikonov\HISKP\hspace{-1mm}$^,$\GATCHINA\hspace{-1mm},
D.~Novinski\GATCHINA\hspace{-1mm},\\
M.~Ostrick\PI\hspace{-1mm}, H.~van~Pee\HISKP\hspace{-1mm},
A.V.~Sarantsev\HISKP\hspace{-1mm}$^,$\GATCHINA\hspace{-1mm},
C.~Schmidt\HISKP\hspace{-1mm}, H.~Schmieden\mbox{\PI\hspace{-1mm},}
B.~Schoch\PI\hspace{-1mm},
G.~Suft\ERLANGEN\hspace{-1mm}, V.~Sumachev\GATCHINA\hspace{-1mm},
T.~Szczepanek\HISKP\hspace{-1mm}, U.~Thoma\HISKP\hspace{-1mm},
D.~Walther\HISKP\hspace{-1mm}, and
Chr.~Weinheimer\HISKP \\
(The CB-ELSA Collaboration)}
\address{$^1$ Helmholtz-Institut f\"ur Strahlen- und Kernphysik der
Universit\"at Bonn, Germany}
\address{\GATCHINA Petersburg
Nuclear Physics Institute, Gatchina, Russia}
\address{\ERLANGEN Physikalisches Institut,
Universit\"at Erlangen, Germany}
\address{\PI Physikalisches Institut,
Universit\"at Bonn, Germany}
\address{\KVI KVI, Groningen, Netherlands}
\address{\FSU Department of Physics, Florida State University, USA}
\address{\BOCHUM
Physikalisches Institut, Universit\"at Bochum, Germany}
\address{\BASEL Physikalisches Institut, Universit\"at
Basel, Switzerland}
\address{\GIESSEN Physikalisches Institut, Universit\"at Giessen}
\date{\today}

\begin{abstract}
Evidence is reported for the existence of a parity doublet of
$\Delta$ resonances with total angular momentum $J=3/2$ from
photoproduction of the $p\pi^0\eta$ final state. The two parity
partners $\Delta(1920)P_{33}$ and $\Delta(1940)D_{33}$ make
significant contributions to the reaction. Cascades of resonances
into $\Delta(1232)\eta$, $N(1535)\pi$, and $Na_0(980)$ are clearly
observed.
\end{abstract}

\maketitle

Chiral symmetry and chiral symmetry breaking play key roles for
understanding strong interactions. Chiral symmetry implies that
chiral partners having the same angular momentum $J$ but opposite
parity should be degenerate in mass. The spin-parity partner of the
nucleon with $J^P=1/2^-$, $N(1535)$, is found 600\,MeV above the
nucleon; the mass splitting between the chiral partners $N(1520)$
with $J^P=3/2^-$ and $\Delta(1232)$ with $J^P=3/2^+$ is 288\,MeV.
The large mass gaps are indications that chiral symmetry is broken
spontaneously.

Surprisingly, the high-mass light baryon spectrum exhibits a new
phenomenon, the occurrence of parity doublets where pairs of $N^*$
or $\Delta$ resonances with the same $J$ but opposite parities form
(nearly) mass-degenerate doublets. In 1999, Glozman suggested to
explain the mass degeneracy by assuming that chiral symmetry is
restored in highly excited hadronic states \cite{Glozman:1999tk}.
The concept of chiral symmetry breaking and its restoration was
extended to postulate chiral multiplets \cite{Cohen:2001gb} in which
all four nucleon and $\Delta$ resonances with the same $J$ are
degenerate in mass.  A new and expanding field - reviewed in
\cite{Jaffe:2006jy,Glozman:2007ek} - was created. In a recent paper,
Shifman and Vainshtein pointed out \cite{Shifman:2007xn} that in
highly excited mesons, chiral symmetry seems not to be restored
since some mass gaps in would-be chiral multiplets are of the same
order of magnitude as mass gaps on Regge trajectories per unit of
angular momentum.

The high theoretical interest in chiral multiplets is, however, not
balanced by reliable experimental information. In
\cite{Glozman:2007ek}, seven chiral multiplets (with $J=1/2$ to
$J=15/2$) are listed but none of them comprises four established
states (with 3-star or 4-star rating in the Review of Particle
Properties (RPP) \cite{Yao:2006px}). From the 14 parity doublets
only two consist of a pair of established resonances, one of them
being questioned in \cite{Arndt:2006bf}. Obviously, new studies are
needed to explore the high-mass region of nucleon and $\Delta$
resonances.

This letter reports evidence for a parity doublet of $\Delta$ states
with $J^P=3/2^{\pm}$ at a mass of $\approx 1980$\,MeV from a study
of the reaction
\begin{equation} \label{R1} \gamma p \to  p\pi^0\eta\ \end{equation}
for photon energies from the $p\pi^0\eta$ production threshold up to
3\,GeV. RPP lists two $J=3/2$ states in this mass range, the
positive-parity $\Delta(1920)$ with 3 stars and the negative-parity
$\Delta(1940)$. Neither of them was seen in \cite{Arndt:2006bf}.

Reaction (1) is very well suited to study decays of $\Delta$
resonances into $\Delta(1232)\eta$. The $\eta$ acts as an isospin
filter: resonances decaying into $\Delta(1232)\eta$ must have
isospin $I=3/2$ and belong to the $\Delta$ states. For low photon
energies phase space is limited, and $\Delta(1232)$ and $\eta$
should be in a relative S-wave. We thus may expect a high
sensitivity for baryon resonances with isospin $I=3/2$, spin $J=3/2$
and negative parity. If such resonances decay into $N\pi$, they need
$L=2$ between $N$ and $\pi$; resonances with these quantum numbers
are characterized by $L_{2I,2J}=D_{33}$. $\Delta(1232)$ and $\eta$
in a relative P-wave can be produced from the $L_{2I,2J}=P_{33}$
wave with isospin $I=3/2$, spin $J=3/2$ and positive parity.

The experiment was carried out at the tagged photon beam of the {\bf
EL}ectron {\bf S}tretcher {\bf A}ccelerator ELSA at Bonn
\cite{Hillert:2006yb}, using the Crystal Barrel detector
\cite{Aker:1992ny}. A description of the experiment and data
reconstruction can be found elsewhere
\cite{vanPee:2007,Thoma:2007bm}, a more detailed documentation of
this analysis in \cite{Horn:2008}. The data cover photon energies
from the $p\pi^0\eta$ production threshold to 3.0\,GeV/c$^2$.

Events due to reaction (\ref{R1}) were selected by the following
cuts: five clusters of energy deposits were required in the Crystal
Barrel calorimeter, one proton and four photons. The proton was
identified by matching one cluster to the charged particle emerging
from the liquid H$_2$ target and hitting a three--layer
scintillation fiber detector \cite{Suft:2005} surrounding the liquid
H$_2$ target (length 5\,cm, diameter 3\,cm). The fast coincidence
between a tagged photon and a hit in the scintillation fiber
detector was used as first level trigger. A second level trigger
required at least three clusters of energy deposits in the Crystal
Barrel. This decision needed about 5\,$\mu$s.

\begin{figure}[pb]
\vspace{-4mm}\centering{
\epsfig{file=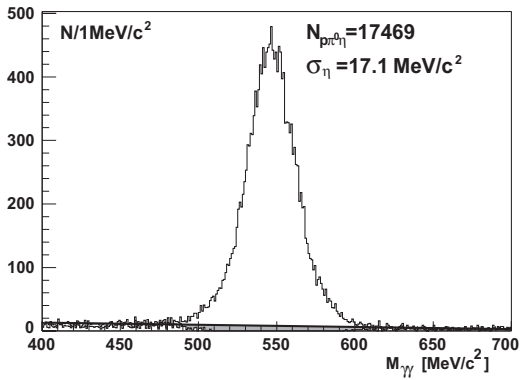,width=0.38\textwidth,height=0.28\textwidth,clip=}}
\vspace{-2mm} \caption{\label{gg_vs_gg}The $\gamma\gamma$ invariant
mass distribution after a kinematic fit to $\rm\gamma p \to p
\pi^0\gamma\gamma$ (with c.l.$>10$\%) and a cut rejecting $\rm\gamma
p \to p \pi^0\pi^0$ events (with c.l.$>1$\%).  }
\end{figure}

These events were subjected to a kinematic fit imposing energy and
momentum conservation. Proton and the four photons were assumed to
be produced in the target center. In a first step, the $\gamma p\to
p\pi^0\gamma\gamma$ hypothesis was tested and events with a
probability (c.l.) exceeding 10\% were retained. In a next step,
events compatible, at a c.l.\,$>$1\%, with the $\gamma p\to p2\pi^0$
hypothesis were rejected. The resulting $\gamma\gamma$ invariant
mass of the photon-pair is shown in Fig.~\ref{gg_vs_gg}. These
events passed a final kinematic fit to the $\gamma p\to p\pi^0\eta$
hypothesis requiring a probability exceeding 1\%. This sample
contains 17469 events due to reaction (\ref{R1}) and 910 background
events  (shaded area in Fig. \ref{gg_vs_gg}). The 910 events closest
in phase space to the events falling into the $\eta$ side bins
(380-440\,MeV/c$^2$; 640-700\,MeV/c$^2$) were subtracted to obtain
the final event sample.

\begin{figure}[pt] \centerline{
\epsfig{file=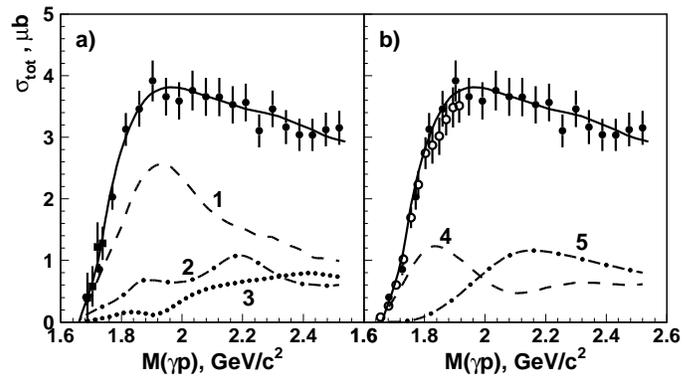,width=0.5\textwidth,clip=}}
\caption{\label{tot_pwa}Total cross sections for $\gamma p \to p
\pi^0\eta$. $\bullet$ this work; (left) black squares
\cite{Nakabayashi:2006ut}, (right) open circles \cite{Ajaka:2008}.
The solid curve represents a PWA fit. (a): Excitation functions. (1)
shows the contribution from the $\Delta(1232)\eta$ intermediate
state, (2) the $N(1535)\pi$, and (3) the $N\,a_{0}(980)$
contribution. (b): Contributions of the $D_{33}$ (4) and $P_{33}$
(5) partial waves.\vspace{-3mm} }
\end{figure}

In Fig.~\ref{tot_pwa}a and b the total cross section is displayed.
The errors include the statistical error and the systematic errors
due to event reconstruction ($\approx 6$\%) and due to the
acceptance correction ($\approx 5$\%) deduced from different
solutions of the partial wave analysis (PWA) described below. A
$\pm15$\% systematic error (not shown) is assigned to the
uncertainty in the photon flux normalization. The solid curve shows
the PWA result. The total cross section reaches a maximum of almost
4\,$\mu$b in the 2\,GeV/c$^2$ region and then de\-creases slowly to
3\,$\mu$b.  Our cross section is compatible with results obtained at
Sendai \cite{Nakabayashi:2006ut} (see Fig.~\ref{tot_pwa}a)  and
GRAAL \cite{Ajaka:2008} (see Fig.~\ref{tot_pwa}b).

\begin{figure*}[pt]
\begin{minipage}[c]{0.66\textwidth}
\begin{tabular}{cc}
\hspace{-3mm}\epsfig{file=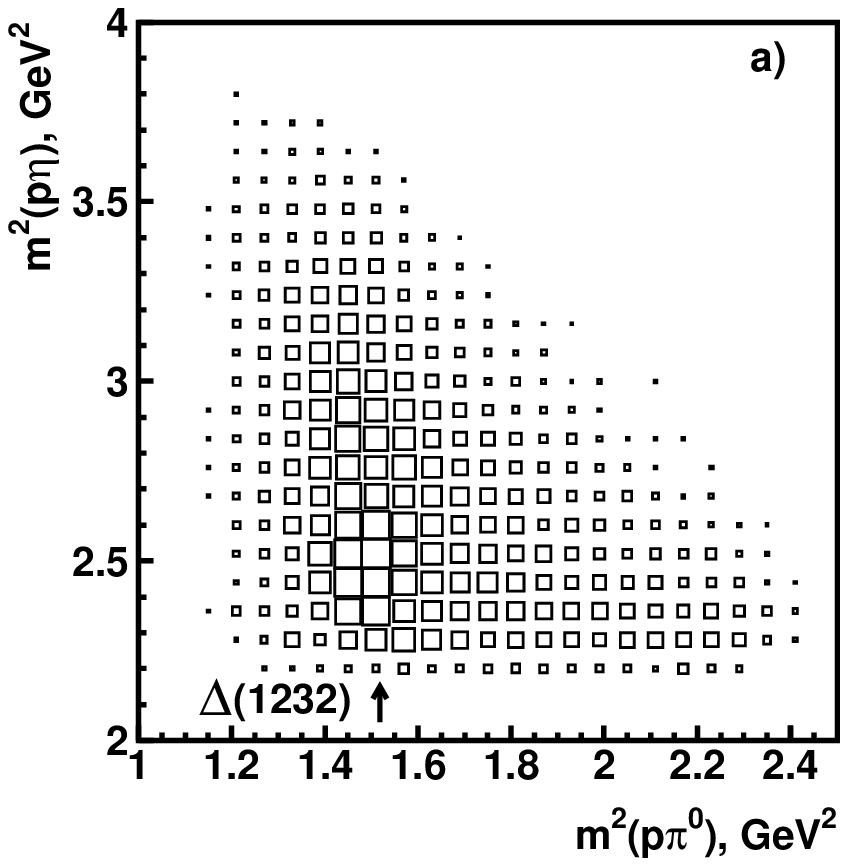,width=0.5\textwidth,height=0.48\textwidth,clip=}&
\hspace{-4mm}\epsfig{file=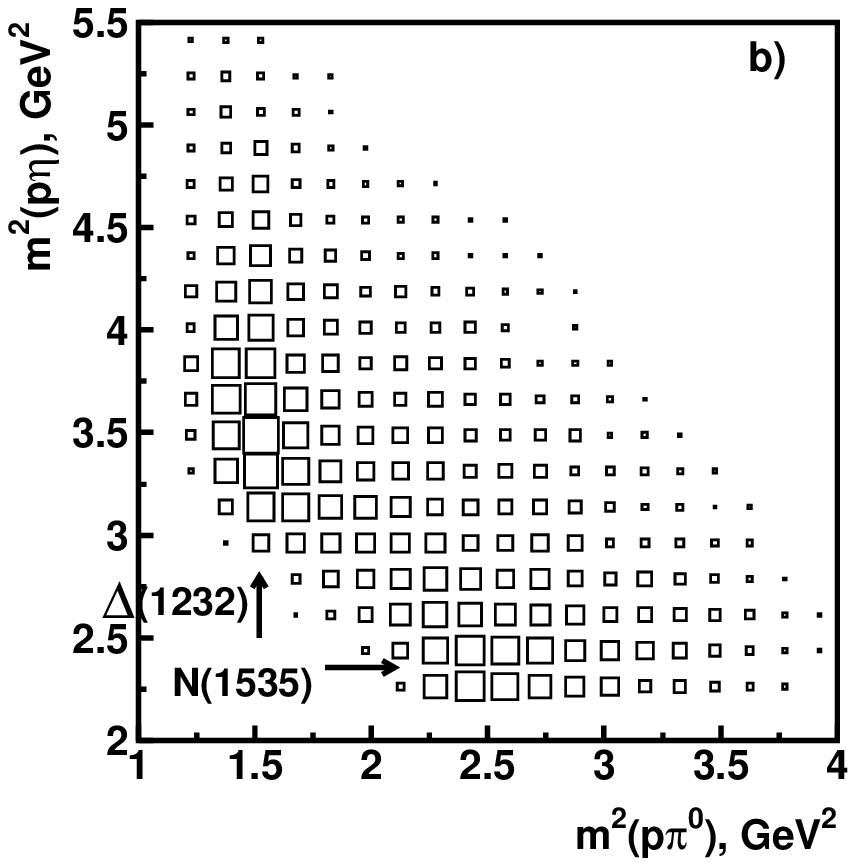,width=0.5\textwidth,height=0.48\textwidth,clip=}
\end{tabular}
\end{minipage}
\hspace{-6mm}\begin{minipage}[c]{0.33\textwidth}
\epsfig{file=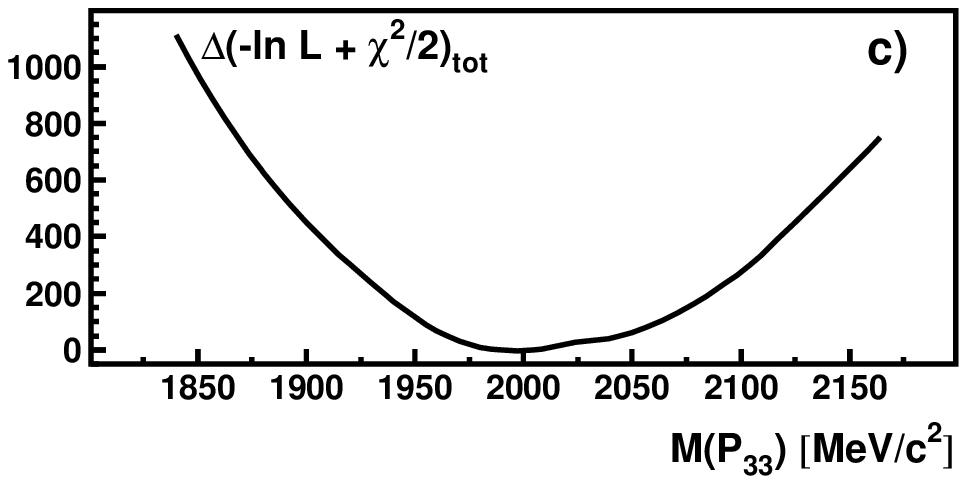,width=1.04\textwidth,height=2.8cm,clip=}\\
\epsfig{file=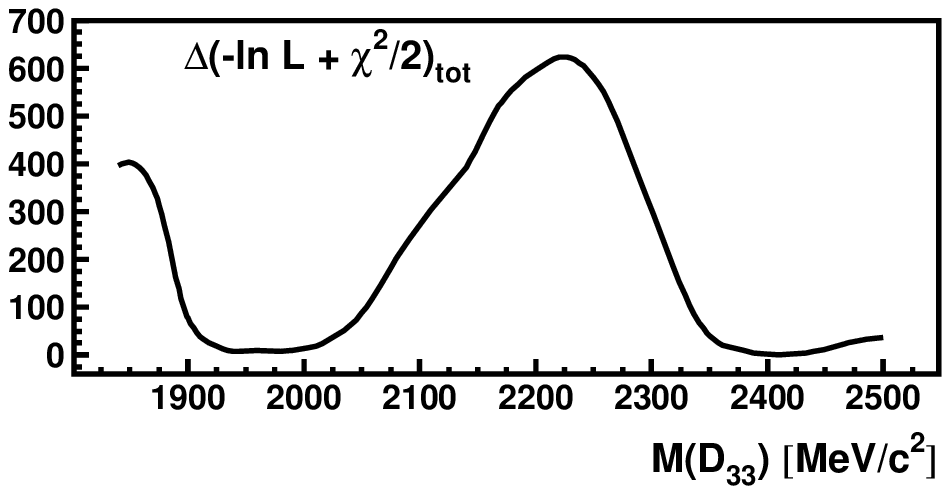,width=1.04\textwidth,height=2.8cm,clip=}
\end{minipage}
\vspace{-2mm}\caption{\label{Dalitz}Dalitz plot for the reaction
$\gamma p\to p\,\pi^0\eta$ for $E_{\gamma}<1.9$\,GeV (a) and
$E_{\gamma}>1.9$\,GeV (b). $\Delta(1232)$ is seen in both Dalitz
plots; $N(1535)$ is visible only for high photon energies even
though the $N(1535)\pi$ production threshold ($\sim1.0$\,GeV) is
lower than the $\Delta(1232)\eta$ production threshold
($\sim1.2$\,GeV). c) Total likelihood as a function of the mass of a
$\Delta P_{33}$ (top) or $\Delta D_{33}$ (bottom) resonance. The
width is fixed to 350\,MeV/c$^2$. \vspace{2mm}}
\centering{
\epsfig{file=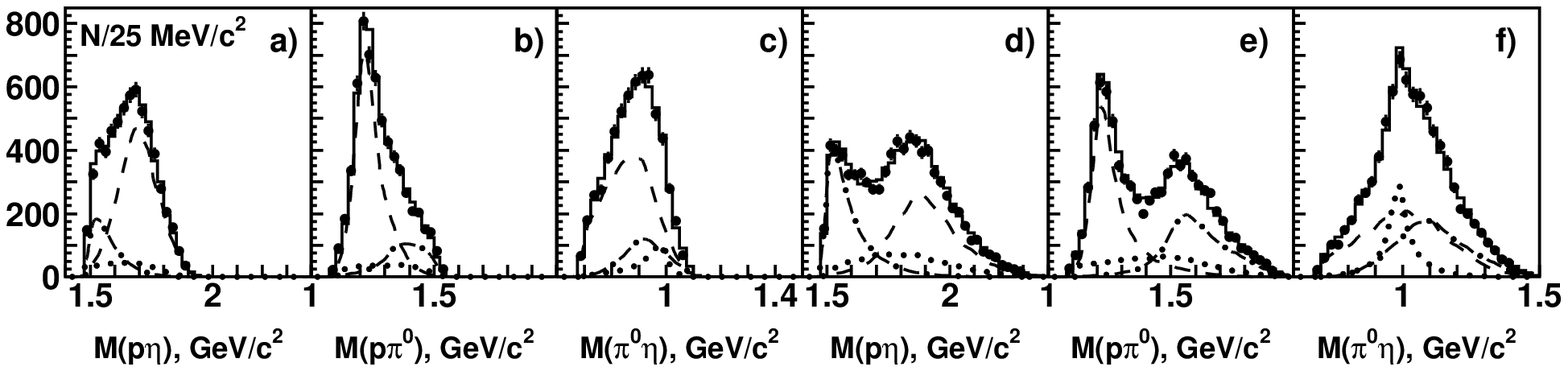,width=\textwidth,height=0.18\textheight,clip=}}
\vspace{-6mm}\caption{\label{dcs_tot}  Mass distributions for
$E_{\gamma}<1.9$\,GeV (a-c) and $E_{\gamma}>2.1$\,GeV (d-f). The
distributions are not corrected for the detector acceptance or
undetected decay modes. The solid curves show the result of the
partial wave analysis. The dashed lines represent the
$\Delta(1232)\eta$ intermediate state, the dashed-dotted lines
$N(1535)\pi$, the dotted lines $pa_0(980)$.\vspace{-3mm}}
\end{figure*}

A few leading contributions for the low- and high-mass range are
visible in the Dalitz plots (Fig.~\ref{Dalitz}). The low-mass data
are dominated by the $\Delta(1232)\eta$ intermediate state and there
is, in agreement with GRAAL, no visible $N(1535)\pi$ contribution.
At large masses, $N(1535)$ can be seen in its $p\eta$ decay. The
mass projection and the partial wave analysis reveal the existence
of $p\,a_0(980)$ as third contribution.

Fig. \ref{dcs_tot} shows the $p\eta$ (a,d), $p\pi^0$ (b,e), and
$\pi^0\eta$ (c,f) mass distributions for $W<1.9$\,GeV/c$^2$ and
$W>$2.1\,GeV/c$^2$, respectively. In Figs.~\ref{dcs_tot}b,e large
$\Delta(1232)$ contributions are observed. Fig.~\ref{dcs_tot}d
reveals $N(1535)\pi$ with $N(1535)$ decaying into $N\eta$; in
Fig.~\ref{dcs_tot}f, $pa_0(980)$ - with $a_0(980)\to\pi^0\eta$ -
becomes visible. In the 1.9 to 2.1\,GeV/c$^2$ mass region (not
shown), $\Delta(1232)\eta$ is still significant, and a small $p\eta$
threshold enhancement due to $N(1535)\pi$ appears.

These findings are confirmed in a partial wave analysis (PWA). The
$P_{33}$ and $D_{33}$ partial waves discussed here (and three
further important partial waves $S_{11}$, $P_{11}$, and $P_{13}$)
are described by K-matrices \cite{Anisovich:2007bq}. The
corresponding elastic $\pi N$ scattering amplitudes from
\cite{Arndt:2006bf} were added to the combined analysis for
invariant masses up to 2.4\,GeV/c$^2$ (2.25\,GeV/c$^2$ for
$P_{33}$). Thus our results contain the full information leading to
the results presented in \cite{Arndt:2006bf}. The fit to the elastic
amplitudes can be viewed in \cite{Horn:2008} (Fig. 8). However, our
analysis is constrained by photo-production data.

In addition to the data on $\gamma p\to p\pi^0\eta$ presented here,
data on photoproduction of single pions, $2\pi^0$, $\eta$, and of
$\Lambda$ and $\Sigma$ hyperons are included. References to the
data, an outline of the PWA method and the definition of likelihood
contributions can be found in \cite{Anisovich:2007bq}. Included here
are new data on the beam asymmetry for $\gamma p\to p\pi^0\eta$
\cite{Gutz:2008zz}. The different data enter with weights $w_i$. The
weights range from $w=1$ for high-statistics data which we want to
be described approximately to $w=30$ for low-statistics data which
we insist to be described well by the fit. The mean weight is
$\overline w=4.2$. For this data and those from \cite{Gutz:2008zz},
$w_{\gamma p\to p\pi^0\eta}=10$ is chosen. Too small a weight leads
to a bad description of this data, a very large weight deteriorates
the description of other data.

The data on reaction (1) presented here, and those on $\gamma p\to
p\pi^0\pi^0$, are fit using an event-based maximum likelihood method
taking into account all correlations between variables in the
5-dimensional phase space. Minimized is the weighted sum of the
(negative) logarithmic likelihood from 3-body final states and the
$\chi^2/2$ contribution from other data.

The excitation functions for $\Delta(1232)\eta$, $N(1535)\pi$, and
$pa_0(980)$ deduced from the PWA, are shown in Fig. \ref{tot_pwa}a.
$\Delta(1232)\eta$ makes the most significant contribution; close to
the threshold, it dominates the reaction. It reaches a maximum just
below 2\,GeV/c$^2$. $N(1535)\pi$ exhibits a structure at
1.9\,GeV/c$^2$ due to the $N(1880)$ (with $P_{11}$ quantum numbers,
formerly called $N(1840)$ \cite{Sarantsev:2005tg}) and
$\Delta(1940)D_{33}$, and a second bump at 2.2\,GeV/c$^2$ which we
interpret as nucleon resonance with $P_{13}$ quantum numbers. With
increasing mass, $pa_0(980)$ gains a notable intensity. In Fig.
\ref{tot_pwa}b, contributions of individual partial waves are shown.
The $D_{33}$ wave rises quickly above the threshold, mostly due to
$\Delta(1940)D_{33}$. The $P_{33}$ partial wave shows no significant
features.

\begin{table*}[pt] \caption{\label{Table:p33_3}
Properties of the $\Delta(1920)P_{33}$ and $\Delta(1940)D_{33}$
resonances.  Masses and widths are given in MeV/c$^2$, branching
ratios in \%. The branching ratios are corrected using
Clebsch-Gordan coefficients and for unseen decay modes of
final-state mesons ($\pi^0$ and $\eta$). The helicity couplings are
in GeV$^{-1/2}$.\vspace{-2mm}}
\begin{center}
\renewcommand{\arraystretch}{1.3}
\begin{tabular}{ccccccccccc} \hline\hline
&$M_{pole}$&$\Gamma_{pole}$& $M_{BW}$&$\Gamma^{BW}_{tot}$&${\rm
Br}_{N\pi}$&${\rm Br}_{\Delta(1232)\eta}$&${\rm
Br}_{N(1535)\pi}$&${\rm Br}_{Na_0(980)}$&$A_{1/2}$&$A_{3/2}$\\
$\Delta(1920)P_{33}$& $1980^{+25}_{-45}$
&$350^{+35}_{-55}$&$1990\pm35$&$375\pm50$&$15\pm8$&$10\pm5$&$6\pm 4$&$4\pm2$&$22\pm8$&$42\pm12$ \\
$\Delta(1940)D_{33}$&$1985\pm30$&$390\pm50$&$1990\pm40$&$410\pm70$
&$9\pm4$&$5\pm2$&$2\pm1$&$2\pm1$&$160\pm40$&$110\pm30$\\
&\phantom{rrrrr}&\phantom{rrrrrrrrrrrr}&
\phantom{rrrrrrrrrrrr}&\phantom{rrrrrrrrrrr}&\phantom{rrrrrrrrr}&
\phantom{rrrrrrr}&\phantom{rrrrrrr}&\phantom{rrrrrrr}&\phantom{rrrrrrr}&\phantom{rrrrrrr}\vspace{-5mm}\\
\hline\hline
\end{tabular}\vspace{-5mm} \end{center}\renewcommand{\arraystretch}{1.0}
\end{table*}

The fit requires contributions from eight resonances; two of them do
not couple to $\Delta(1232)\eta$ and are assigned to nucleon
excitations. Most pole positions agree with previously reported
values. The $N(1880)$ is observed at $M-\frac{i}{2}\Gamma =
1880-i110$\,MeV/c$^2$. It decays to $N(1535)\pi$ and contributes
$\sim12$\% to reaction (\ref{R1}). $N(2200)P_{13}$ decays mainly
into $N\,a_0(980)$. These two resonances were already required in an
analysis of $\gamma p\to N\pi$, $\Lambda K^+$ and $\Sigma K^+$
\cite{Sarantsev:2005tg}. Here, they are found to contribute to
$p\pi^0\eta$. However, due to limited statistics and without
measurements of polarization variables covering the region above
2\,GeV/c$^2$, quantum number assignments above 2\,GeV/c$^2$ have to
be taken with some precaution. $\Delta(1905)F_{35}$ contributes a
small but significant fraction. It is observed with a pole at
$1920-i145$\,MeV/c$^2$. The other states belong to the $P_{33}$ and
$D_{33}$ wave which are discussed next.

The $P_{33}$ and $D_{33}$ are described by K-matrices with 3-pole
6-channel ($\pi N$, $\Delta(1232)\pi$ ($P,F$-waves),
$\Delta(1232)\eta$, $N(1535)\pi$, $N\rho$) K-matrix  in the $P_{33}$
wave and 2-pole 6-channel ($\pi N$, $\Delta(1232)\pi$ ($S,D$-waves),
$\Delta(1232)\eta$, $N(1535)\pi$, $N\rho$) K-matrix in the $D_{33}$
wave. The statistical evidence for the two suggested states (see
Table \ref{Table:p33_3}) was estimated by removing them individually
from the fit. The masses, widths and branching ratios of the
Breit-Wigner states given in Table \ref{Table:p33_3} are calculated
from the residues of the K-matrix poles (see \cite{Thoma:2007bm} for
details). They are determined from the range of values obtained in a
large number of fits. The pole position of the second $P_{33}$ state
(above $\Delta(1232)$) is found at $1510^{+20}_{-50}-i\,115\pm
20$\,MeV/c$^2$. $\Delta(1700)D_{33}$ is observed in decays to
$\Delta(1232)\eta$ and $N(1535)\pi$ with a pole at
$1650-i\,160$\,MeV/c$^2$.

To this solution we added, one by one, Breit-Wigner amplitudes in
different partial waves. The fit improves notably by adding a third
$D_{33}$ state; the statistical significance is $4$ standard
deviations for this data and more than $10\sigma$ for the overall
fit. Its mass optimizes at about 2.36\,GeV/c$^2$ and at a width in
the 400-600\,MeV/c$^2$ range. However, its mass is close to the end
of the available phase space. An upper bound for its mass is not
well defined and we do not claim that this state exists. With this
state included, the $D_{33}$ elastic scattering amplitude can be
fitted up to 2.4\,GeV/c$^2$ invariant mass. By adding resonances
with other quantum numbers we did not find a significant
improvement. In all waves the fit produced a resonance with mass at
the end of phase space and a very broad width. The only exception
was the $D_{35}$ wave which optimized with $M\approx
2150$\,MeV/c$^2$, $\Gamma\approx 160$\,MeV/c$^2$, and a 1\%
fractional contribution. However the statistical evidence for this
state was marginal, $\approx 2\sigma$. No evidence was found for
$\gamma p\to N(1520)\eta$ or $\gamma p\to \Delta(1600)\eta$.

Finally, we performed mass scans for the $P_{33}$ and $D_{33}$ waves
by replacing the highest-mass K-matrix pole in the solution above by
a relativistic Breit-Wigner amplitude. Then the masses of the
Breit-Wigner amplitudes were scanned in the region 1.85-2.4\,GeV.
The result of the mass scan is shown in Fig.~\ref{Dalitz}c. In the
$P_{33}$ scan, the likelihood has a clear minimum around
1980-1990\,MeV/c$^2$, while the $D_{33}$ scan reveals two minima at
1940-2000 and 2400\,MeV/c$^2$. The clear minima support our claim
that both, $\Delta(1920)P_{33}$ and $\Delta(1940)D_{33}$, are
observed in the reaction $\gamma p\to p\pi^0\eta$. Further details
of the analysis can be found elsewhere \cite{Horn:2008}.

Summarizing, we have reported evidence for a $J=3/2$ parity doublet
of $\Delta$ resonances at $\approx$ 1980\,MeV/c$^2$ from $\gamma p
\to p\pi^0\eta$. The parity doublet is compatible with the
conjectured restoration of chiral symmetry at high baryon excitation
energies. Parity doublets are not expected in quark models even
though our findings are - within experimental errors - not
incompatible with quark model predictions and certainly compatible
with predictions based on QCD/AdS \cite{Klempt:2008}.

We would like to thank the technical staff of ELSA and of the
participating institutions for their invaluable contributions to the
success of the experiment. We acknowledge financial support from the
Deutsche Forschungsgemeinschaft (DFG-TR16) and the Schweizerische
Nationalfond. The collaboration with St. Petersburg received funds
from DFG and RFBR. 
\end{document}